# Inverted Gabor holography principle for tailoring arbitrary shaped three-dimensional beams

Tatiana Latychevskaia* and Hans-Werner Fink
Physics Department, University of Zurich
Winterthurerstrasse 190, 8057 Zurich, Switzerland
*Corresponding author: tatiana@physik.uzh.ch

**ABSTRACT**
It is well known that by modifying the wavefront in a certain manner, the light intensity can be turned into a certain shape. However, all known light modulation techniques allow for limited light modifications only: focusing within a restricted region in space, shaping into a certain class of parametric curves along the optical axis or bending described by a quadratic-dependent deflection as in the case of Airy beams. We show a general case of classical light wavefront shaping that allows for intensity and phase redistribution into an arbitrary profile including pre-determined switching-off of the intensity. To create an arbitrary three-dimensional path of intensity, we represent the path as a sequence of closely packed individual point-like absorbers and simulate the in-line hologram of the created object set; when such a hologram is contrast inverted, thus giving rise to a diffractor, it creates the pre-determined three-dimensional path of intensity behind the diffractor under illumination. The crucial parameter for a smooth optical path is the sampling of the predetermined curves, which is given by the lateral and axial resolution of the optical system. We provide both, simulated and experimental results to demonstrate the power of this novel method.

## INTRODUCTION

A classical light beam propagates along a straight line and does not bend unless in a medium of variable refractive index. It is well known that by modifying the wavefront in a certain manner, the light intensity can be turned into a certain shape. Examples are optical lenses or Fresnel Zone Plates for focusing an incident wave to a point at the focal plane. Another example are Airy beams[1] created by modifying the phase distribution of the wavefront into an Airy function resulting in a bending of the light intensity while propagating. A further example is holography, where the phase of the wavefront passing through a hologram is changed to mimic the object wavefront, thus providing the illusion that the original object is present in space[2]. However, all these known techniques allow for limited light modifications: focusing within a limited region in space[2], shaping into a certain class of parametric curves[3-8] along the optical axis or bending described by a quadratic-dependent deflection as in the case of Airy beams[1]. A detailed overview of the existing light modulation methods and their limits is provided by Piestun and Shamir[9]. In the presented here work we show a general case of classical light wavefront shaping that allows for intensity and phase redistribution into an arbitrary shape including pre-determined switching-off of the intensity. To demonstrate our technique we reshape the intensity and the phase of the beam into various three-dimensional shapes also adding controllable switching off the intensity. Both, simulations and experiments are presented to demonstrate the power of this novel tool.

Even a simple transparency can be employed to modify the amplitude of a wavefront as has been demonstrated by Dennis Gabor in his first holographic experiments[2,10]. With the availability of spatial light modulators (SLMs) it has become possible to change amplitude and phase of the incident wavefront in a 'plug and play' manner. SLM can modify the wavefront for various exciting experiments: for imaging through turbid layers[11], creating Airy beams[1,12] or optical vortices[13] etc.

In 1979 Berry and Balazs have theoretically predicted the existence of beams which are bending while propagating[1]. These special beams are described by Airy functions, hence the name Airy beams. Though Berry and Balazs addressed beams of particles of certain mass, the first realization of Airy beams was demonstrated with photons by Siviloglu et al in 2007[12] who employed a SLM to modify the phase distribution of the incident wavefront accordingly.

In principle, Airy beams can be considered as a classical beam with amplitude and phase of the wavefront that are modulated in such a way that it creates an illusion of a bending collimated beam. The propagation of Airy beams is described in a manner no different than classical wavefront propagation, namely, by Fresnel integrals based on Huygens principle. This makes one wonder whether it might be possible to modify the wavefront in such a manner that the beam undertakes a certain at will pre-defined arbitrary intensity distribution while it is propagating. Previous works studies have attempted to modify the intensity of a coherent beam in a predefined manner[8,14]. Whyte et al tried implementing the three-dimensional Gerchberg-Saxton algorithm to obtain a complex-valued distribution on the Ewald sphere in *k*-space related to a phase-modulating pattern which should create a shaped intensity modulation. However, the experimental results only weakly resembled the expected light distribution. Rodrigo et al experimentally realized so-called parametric spiral beams[15] using a dedicated SLM modulating only the phase. In our work we show how light of arbitrary three-dimensional intensity distribution can be created using a relatively simple holographic approach.

## RESULTS
### The principle
The overall idea behind creating a modulation of the intensity of light into an arbitrary curve can be explained by considering the following example. A Fresnel Zone Plate (FZP) can be created by the interference pattern between a plane and a spherical wave. Alternatively, it can be created by an in-line hologram of a point-like absorbing (opaque) particle. Such hologram, when reconstructed, forms the image of the point-like opaque particle, as shown in Fig. 1(a). This type of holograms with focusing properties are called holographic optical elements[16]. Next, an interesting observation can be made: When the contrast of such hologram is inverted it becomes a FZP: the wavefront behind the inverted contrast hologram converges towards the point-like intensity distribution which is the focal point of the FZP, as illustrated in Fig. 1(b). Thus, a hologram of an opaque point-like object when inverted creates under reconstruction a point-like region of a higher intensity.

The example illustrated in Fig. 1 can be also explained in terms of a transmission function and a holographic equation. The transmission function in the plane $(x, y)$ where a point-like absorbing object located at $(x = x', y = y')$ is described by

$$t(x, y) = 1 - o(x, y) \tag{1}$$

where $t(x, y) = 1$ (full transmittance) everywhere except for the region where the point-like absorbing object is located, and where $t(x = x', y = y') = 0$. Thus, $o(x, y)$ is a function which has values equal 0 everywhere except for the region $(x = x', y = y')$ where the point-like absorbing object is located, and where $o(x = x', y = y') = 1$. After a plane wave of unit amplitude passes the $(x, y)$-plane, its wavefront is described by $t(x, y)$. At some distance behind the $(x, y)$ - plane, in the $(X, Y)$ - plane, the distribution of the wavefront is described by

$$U(X,Y) = L[1 - o(x, y)] = 1 - O(X, Y), \tag{2}$$

where $L$ is an operator for forward propagation (as described later by Eqs. 6 – 7). The distribution of the formed inline hologram is described as:

$$H(X,Y) = |U(X,Y)|^2 = 1 - O(X,Y) - O^*(X,Y) + |O(X,Y)|^2. \tag{3}$$

Without any object present in the $(x, y)$-plane, the transmission function is $t(x, y) \equiv 1$ and the formed hologram is $H(X,Y) \equiv 1$. The reference wave is the plane wave $R(X,Y) = 1$.

During the reconstruction of the hologram $H(X,Y)$ given by Eq. 3, the hologram is illuminated with the reference wave $R(X,Y) = 1$ and the wavefront is back propagated from the $(X,Y)$ - plane to the $(x, y)$ - plane:

$$L^{-1}[R(X,Y)H(X,Y)] = L^{-1}\left[1 - O(X,Y) - O^*(X,Y) + |O(X,Y)|^2\right] =$$
$$= 1 - o(x, y) + L^{-1}\left[-O^*(X,Y) + |O(X,Y)|^2\right]. \tag{3}$$

where the last two terms represent the twin image and the typically neglected object term respectively. The first two terms constitute the reconstructed transmission function $t(x, y) = 1 - o(x, y)$ as defined by Eq. 1. It has values equal to 1 everywhere except for the position of the absorbing object at $(x=x', y=y')$, where it has values equal 0. Thus, the reconstructed intensity distribution is everywhere bright except for a point-like dark spot at $(x=x', y=y')$.

The contrast inverted hologram is described as:

$$1 - H(X,Y) = O(X,Y) + O^*(X,Y) - |O(X,Y)|^2. \qquad (4)$$

During the reconstruction of the inverted hologram, it is illuminated with the reference wave $R(X,Y) = 1$ and the wavefront is back propagated from the $(X,Y)$-plane to the $(x,y)$-plane:

$$L^{-1}[R(X,Y)(1-H(X,Y))] = L^{-1}[O(X,Y) + O^*(X,Y) - |O(X,Y)|^2] =$$
$$= o(x,y) + L^{-1}[O^*(X,Y) - |O(X,Y)|^2], \qquad (5)$$

where the last two terms again represent the twin image and the typically neglected object term. The first term $o(x,y)$ is defined above in Eq. 1 and is a function which has values equal to 0 everywhere except for the region $(x=x', y=y')$ where the point-like absorbing object is located, and where $o(x=x', y=y') = 1$. Thus, the reconstructed intensity distribution given by the squared amplitude of the result in Eq. 5 is 0 everywhere except for a point-like bright spot at $(x=x', y=y')$.

The images of the holograms shown in Fig. 1 are not just artistic creations, these are images obtained by simulations: a hologram of a point-like absorber and the related inverted hologram.

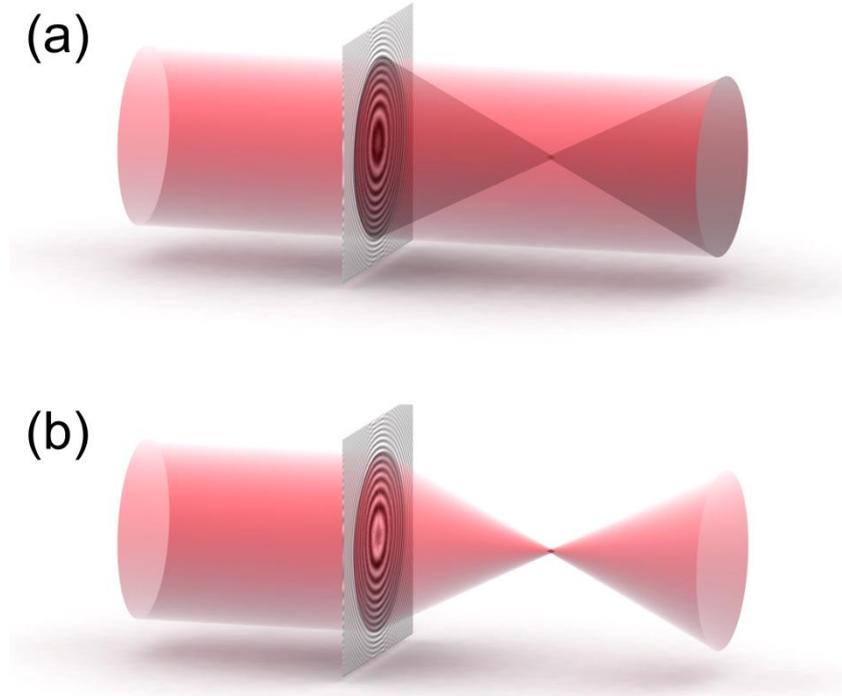

Figure 1. Diffraction of a plane wave on a holographic optical element. (a) An image of a point-like opaque particle is reconstructed behind an in-line hologram of a point-like opaque particle. (b) A point-like intensity distribution is reconstructed behind an inverted in-line hologram of a point-like opaque particle.

**Creating the diffractor**

A hologram is capable of storing three-dimensional information. A good example is holographic particle tracking[17-22], where a single hologram is sufficient to store information about the three-dimensional distribution of particles. During the reconstruction of such a hologram, each of the particles is reconstructed at its $(x, y, z)$ position as a point-like opaque object. Should those particles be aligned along some three-dimensional curve, their point-like opaque reconstructions will appear along that curve. Moreover, when the contrast of the hologram is inverted, their reconstructions will appear as spots of higher intensity along that curve. Using this principle of focusing towards a point, an arbitrary path along the optical axis can be built as a set of point-like particles.

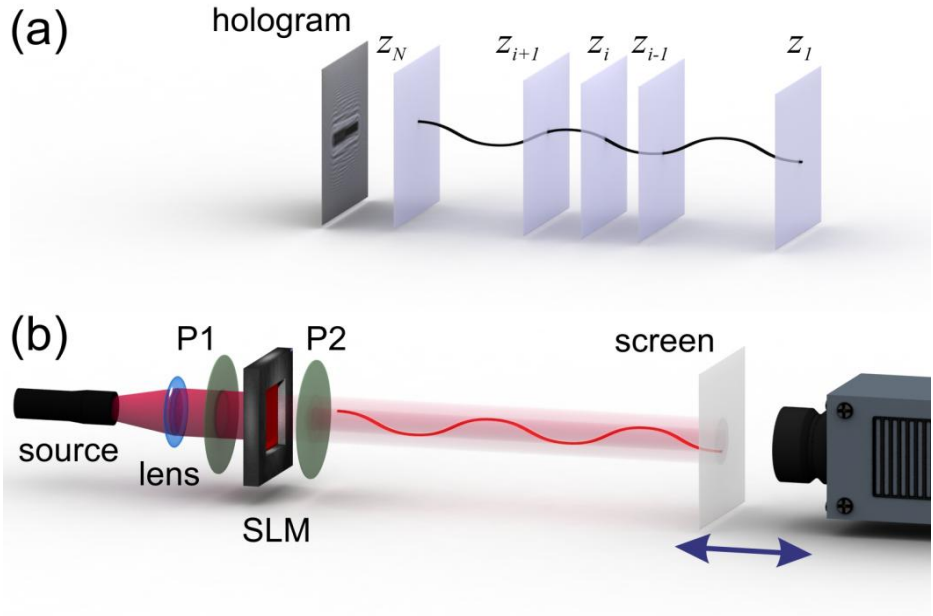

Figure 2. Shaping the intensity of light. (a) A continuous path in the form of a cosine-like curve is sampled in planes at $z_i$, $i = 1...N$, providing a distribution of transmission functions $t_i(x_i, y_i, z_i)$. The hologram $H(X,Y)$ is simulated by propagating an optical plane wave through all planes down to the hologram plane $(X,Y)$. (b) Laser light is collimated by a lens system and directed towards the SLM containing the diffractor distribution $D(X,Y)$. By adjusting the polarizers P1 and P2, a high contrast image is obtained behind the P1-SLM-P2 assembly. The wavefront following the P1-SLM-P2 assembly is shaped into an intense pre-defied path. A semi-transparent screen together with a CCD camera is moved along the optical axis to acquire the intensity distribution at different planes along the path.

Figure 2(a) shows a cosine-like path, which is sampled at planes orthogonal to the optical axis at different distances $z_i$, $i = 1...N$. For a smooth appearance of the optical path, the individual absorbers must be as close to each other as possible. For two absorbers in adjacent planes, the lateral distance in $(x, y)$ plane must not exceed the lateral resolution and the axial distance in $z$-direction must not exceed the axial resolution, see Methods. At each sampling plane, the cross-section of the plane with the path gives rise to a set of point-like objects in that plane. The transmission functions $t_i(x_i, y_i, z_i)$ are assigned to each plane so that $t_i(x_i, y_i, z_i) = 1$ everywhere except at the positions of the point-like objects where $t_i(x_i, y_i, z_i) = 0$. The propagated complex-valued wavefront distribution $U_{i+1}(x_{i+1}, y_{i+1}, z_{i+1})$ at plane $z_{i+1}$ is calculated by using angular spectrum method[23-24]:

$$U_{i+1}(x_{i+1}, y_{i+1}, z_{i+1}) = \text{FT}^{-1}\left\{\text{FT}\left[t_i(x_i, y_i, z_i) U_i(x_i, y_i, z_i)\right] \exp\left[\frac{2\pi i(z_{i+1} - z_i)}{\lambda}\sqrt{1 - (\lambda f_x)^2 - (\lambda f_y)^2}\right]\right\}, \quad (6)$$

where $t_i(x_i, y_i, z_i)$ is the transmission function at plane $z_i$, FT and FT$^{-1}$ are the two-dimensional Fourier and inverse Fourier transforms, respectively, $(f_x, f_y)$ are the spatial frequencies, and $\lambda$ is the wavelength. The contribution of the constant term $\exp\left[\frac{2\pi i(z_{i+1} - z_i)}{\lambda}\right]$ in Eq. 6 can be neglected. Thus, the propagated complex-valued wavefront distribution $U_i(x_{i+1}, y_{i+1}, z_{i+1})$ at plane $z_{i+1}$ is calculated by:

$$U_{i+1}(x_{i+1}, y_{i+1}, z_{i+1}) = $$
$$= \exp\left(-\frac{2\pi i(z_{i+1} - z_i)}{\lambda}\right) \text{FT}^{-1}\left\{\text{FT}\left[t_i(x_i, y_i, z_i)U_i(x_i, y_i, z_i)\right]\exp\left[\frac{2\pi i(z_{i+1} - z_i)}{\lambda}\sqrt{1 - (\lambda f_x)^2 - (\lambda f_y)^2}\right]\right\}. \quad (7)$$

The hologram $H(X,Y)$ is simulated by propagating a plane wave through the sequence of planes with transmission functions $t_i(x_i, y_i, z_i)$, further propagation to the hologram plane $(X,Y)$ and taking the square of the absolute value of the wavefront distribution. The values of hologram distribution $H(X,Y)$ can exceed 1. The diffractor is described by a transmission function, and therefore its values cannot exceed 1. Therefore, the diffractor $D(X,Y)$ is obtained by scaling the hologram amplitude values to the maximum of 1, giving $H'(X,Y)$ and inverting the result by calculating $D(X,Y) = 1 - H'(X,Y)$. It is worth noting that a diffractor cannot be simulated just by summing up light sources along the path, as for example proposed by Rodrigo et al[8]. The intensity behind adjacent sources aligned along the optical axis, will increase in geometrical progression along the path and lead to an extremum in intensity in one direction. As a result, the obtained hologram will not reconstruct such a light extremum as a light trajectory. The approach of summing up intensities from point sources might only be used when the sources are not positioned behind each other along the path in the direction of optical axis, but are distributed along a certain three-dimensional curve[8,15]. Such approach is thus limited to certain class of curves only. In our method, while the path is built up by absorbers, no accumulation of signal occurs along the path, and the corresponding recorded hologram will reconstruct the three-dimensional path with a constant transmission value of 0. To obtain a path of high light intensity, the hologram just needs to be inverted. When illuminating the diffractor $D(X,Y)$ with a plane wave, the intensity distribution behind the diffractor will converge to intense spots at the positions of the point-like objects, thus re-creating the shape of the path, as illustrated in Fig. 2(b). In this manner, arbitrary intensity distributions in space can be created at will provided the length of the path does not exceed the coherence length of the light source.

The numerical reconstruction of the optical wavefront behind the diffractor using the angular spectrum method involves the following calculation[24]:

$$U_i(x_i, y_i, z_i) = \text{FT}^{-1}\left\{\text{FT}\left[D(X,Y)\right]\exp\left[-\frac{2\pi i z_i}{\lambda}\sqrt{1 - (\lambda f_x)^2 - (\lambda f_y)^2}\right]\right\}. \quad (8)$$

The spherical wave term can be expressed in a digital form to account for discrete pixels:

$$\exp\left[-\frac{2\pi i z_i}{\lambda}\sqrt{1 - (\lambda f_x)^2 - (\lambda f_y)^2}\right] \approx \exp\left[-\frac{2\pi i z_i}{\lambda}\right]\exp\left[\frac{\pi i z_i}{\lambda}\left((\lambda f_x)^2 + (\lambda f_y)^2\right)\right] =$$
$$= \exp\left[-\frac{2\pi i z_i}{\lambda}\right]\exp\left[\frac{\pi i z_i \lambda}{s^2}(m^2 + n^2)\right], \quad (9)$$

where $s \times s$ is the area size, and $m$ and $n$ are the pixel numbers. The first term in Eq. 9 is a constant in a plane perpendicular to the optical axis with constant $z_i$. The modulus of this term is one when the amplitude of the reconstructed object distribution is calculated. The squared amplitude of the complex-valued wavefront behind the diffractor, obtained by Eq. 8, provides the intensity distribution of the wave which is measured in the experiment.

**Experimental setup**
The experimental setup is shown in Fig. 2(b), its components are: a laser of 650 nm wavelength, a spatial light modulator (SLM) LC2002 operating with twisted nematic liquid crystal with 832 × 624 pixels and a pitch of 32 micron, two polarizers to control the contrast of the SLM image, and a movable detecting system consisting of a semi-transparent screen (made of Mylar-like material) followed by a CCD camera (Hamamatsu C4742-95, 10 bit or 1024 grey levels, 1280 × 1024 pixels, pixel size 6.7 × 6.7 μm²). The laser with mounted lenses, SLM and polarizers are the parts of a OptiXplorer Educational Kit by Holoeye. The simulated diffractor patterns are loaded onto the SLM as a purely transmitting image. The intensity distributions were acquired at distances ranging from $z = 100$ to 900 mm from the SLM with a step width of 4 mm. Thus, this optical setup allows for a lateral resolution of 22 μm and an axial resolution of 2.1 mm (see Methods, Eq. 14 and 26, calculated at a distance of $z = 900$ mm). The dark current of the CCD results in a noisy background of about 50 grey levels. A constant background intensity of 50 grey levels was subtracted from the measured intensity distributions.

**Intensity modulation into a split beam**

Figure 3 shows the light intensity modulated in such a way as to form a straight beam eventually splitting into two parallel beams. The simulated diffractor is shown in Fig. 3(a) and the pre-defined path in Fig. 3(b). In the simulations presented here we sampled at planes $z = 100 - 900$ mm using steps of 4 mm. When numerically reconstructed, the intensity distribution is matching the distribution of the initial path. The experimentally measured light distribution reproduces the predefined split shape, see Fig. 3(b). The $(x,z)$-plane distributions shown in Fig. 3(c) also show good agreement between the numerically reconstructed and experimentally observed intensity distributions. The $(x,z)$-plane distributions were obtained by averaging over 3 pixels along $y$-direction, that is as $\left[I(x, y = -1\,\text{px}, z) + I(x, y = 0\,\text{px}, z) + I(x, y = +1\,\text{px}, z)\right]/3$. The $(x,y)$-plane distributions at three different distances are shown in Fig. 3(d).

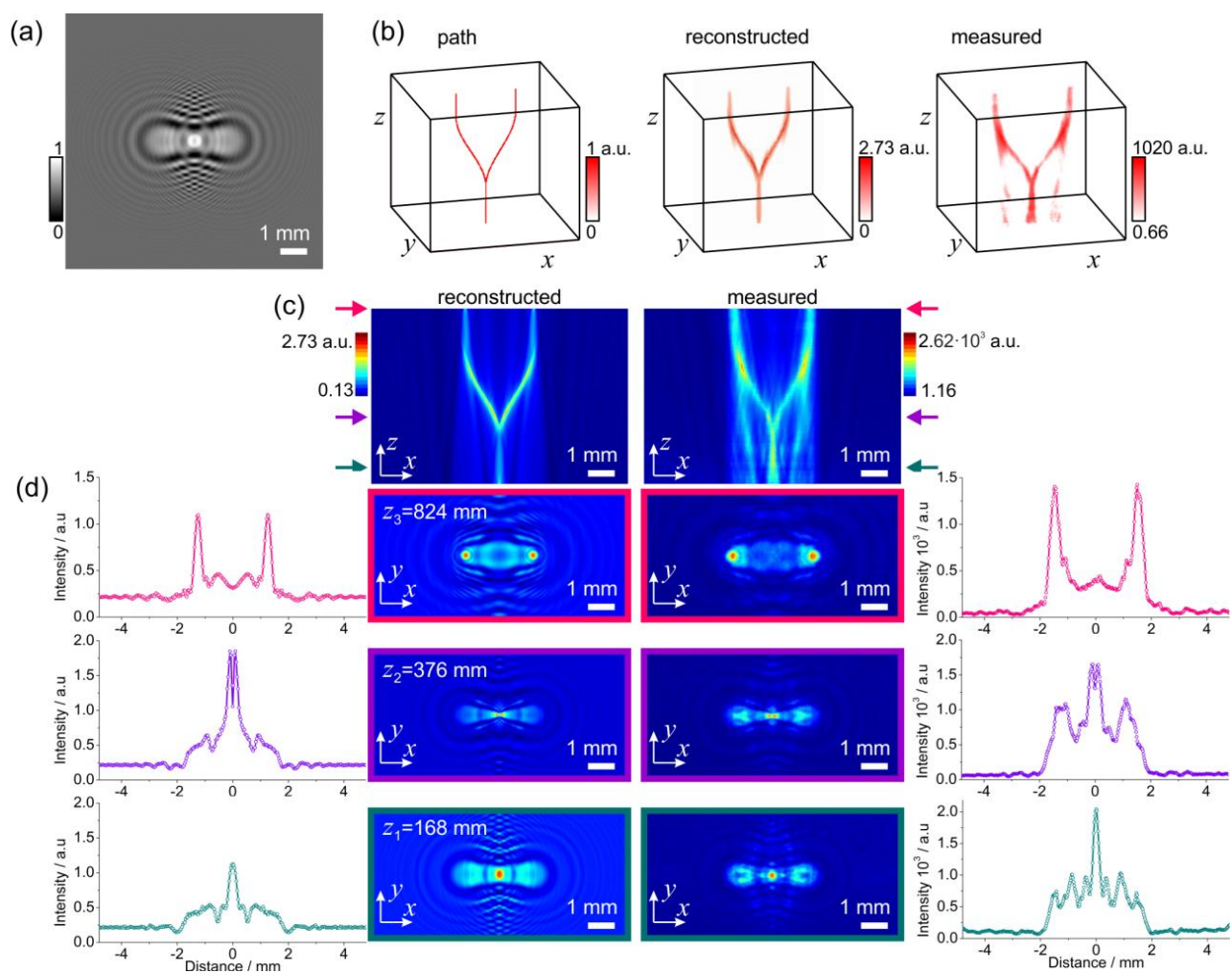

Figure 3. Shaping a split light beam. (a) Central part of 300 × 300 pixels corresponding to 9.6 mm × 9.6 mm of the simulated diffractor. (b) Three-dimensional representation of the light intensity: the desired intensity path used for simulating the diffractor (left), the numerically reconstructed intensity distribution by field propagation behind the diffractor (center), and the experimentally measured intensity distribution behind the diffractor (right). (c) Two-dimensional intensity distributions $I(x, y=0, z)$ numerically reconstructed behind the diffractor (left) and the experimentally measured intensity (right). (d) Intensity distributions at three different $z$-positions: $z_1$ = 168 nm, $z_2$ = 376 nm and $z_3$ = 824 nm indicated with colored arrows in (c). Two-dimensional $I(x, y, z_i)$ at fixed $z_i$-positions intensity distributions obtained by numerical reconstruction (second column) and measured experimentally (third column) behind the diffractor. The related one-dimensional intensity distributions $I(x, y=0, z_i)$ at the corresponding $z_i$-positions are shown in the first and in the fourth column.

**Intensity modulation into a cosine-like bending beam**

The intensity of the light can be shaped into any arbitrary form; an example of a cosine-like shape is demonstrated in Fig. 4. In the simulations presented here we sampled at planes $z = 100 - 900$ mm using steps of 4 mm. The related MATLAB code is provided in the Supplementary Note.

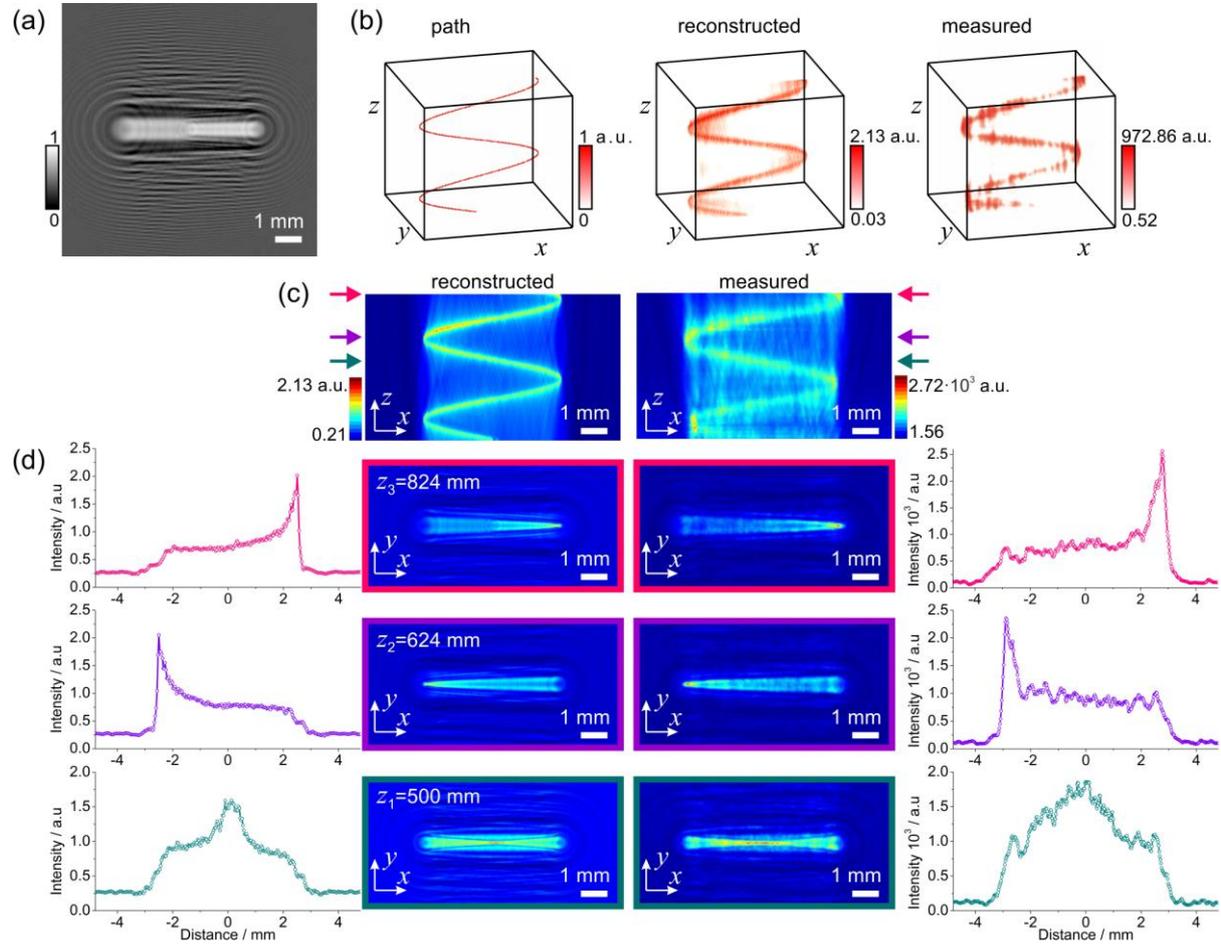

Figure 4. Light bending in cosine-shape. (a) Central part of 300 × 300 pixels corresponding to 9.6 mm × 9.6 mm of the simulated diffractor. (b) Three-dimensional representation of the light intensity: the desired intensity path used for the simulation of the diffractor (left), the numerically reconstructed intensity distribution by field propagation behind the diffractor (center), and the experimentally measured intensity distribution behind the diffractor (right). (c) Two-dimensional intensity distribution $I(x, y = 0, z)$ numerically reconstructed behind the diffractor (left) and the experimentally measured intensity (right). (d) Intensity distributions at three different positions $z_1 = 500$ nm, $z_2 = 624$ nm and $z_3 = 824$ nm indicated with coloured arrows in (c). Two-dimensional $I(x, y, z_i)$ at fixed $z_i$-positions intensity distributions obtained by numerical reconstruction (second column) and measured experimentally (third column) behind the diffractor. The related one-dimensional intensity distributions $I(x, y = 0, z_i)$ at the corresponding $z_i$-positions are shown in the first column and in the fourth column.

**Intensity modulation into a helical-like bending beam and the sampling issue**

Another example, where the intensity was re-shaped into a three-dimensional helix[3,25] is shown in Fig. 5. Here, the central part of the hologram was replaced by a noisy patch to avoid accumulation of the signal in the center of the spiral, and removing part of a hologram does not strongly change the reconstructed object[26].

The diffractor for creation of a helical intensity curve is shown in Fig. 5(a). The helical curve shown in Fig. 5(b) – (c) is sampled with 200 slices in $z$-direction, and is thus created with 200 absorbers (sampling points). The selected helical shape exhibits a period of 70 pixel in $z$-direction and thus runs over 200/70 = 2.86 turns. The radius of helix amounts to 70 pixel and therefore the length of the helical circle in $(x, y)$-plane measures $2\pi \times 70$ pixel = 440 pixel. Thus, it needs at least 440 point-like absorbers per turn for correct sampling in $(x, y)$-plane. Instead, there were only 70 point-like absorbers (sampling points) pro turn. As a result of such under-sampling, the intensity distributions, shown in Fig. 5(b), exhibit some strikes-like appearance.

For a smooth appearance, the distance between the absorbers in the $(x, y)$-plane must not exceed the lateral resolution of the optical system, and the distance between the absorbers in $z$-direction must not exceed the axial resolution.

To improve the appearance of the curve, another helical curve with optimized parameters was created. It has a radius of 110 pixels and thus a length of the circle of $2\pi \times 110$ pixel = 691 pixel. With 2000 sampling points in $z$-direction, 700 point-like absorbers per turn were created. This allows for correct sampling because the number of absorbers per turn (700 absorbers) is higher than the number of points to be sampled (691 pixels). The diffractor was simulated from 2000 slices and its reconstruction obtained at 2000 reconstruction planes shown in Fig. 5(d) exhibits an intensity curve with a better smoothness. Moreover, the experimentally obtained intensity distribution measured at 680 $z$-planes provided in Fig. 5(d) exhibits a smoother appearance when compared with the under-sampled curve shown in Fig. 5(b). This example thus illustrates how correct sampling is important for a smooth appearance of the curve.

From the experimental examples demonstrated in Fig. 3 – 5, it is apparent that the contrast in the experimental images is superior to that in the numerically reconstructed images. The intensity in the vicinity of the path is almost zero. This difference between the experimental and numerically reconstructed distributions can be explained by the fact that the SLM system with polarizers allows for manual contrast adjustment, and polarization is not taken into account into the simulation of the diffractor. The remaining signal around the shaped path is mainly due to the twin image which is intrinsic to in-line holography[27].

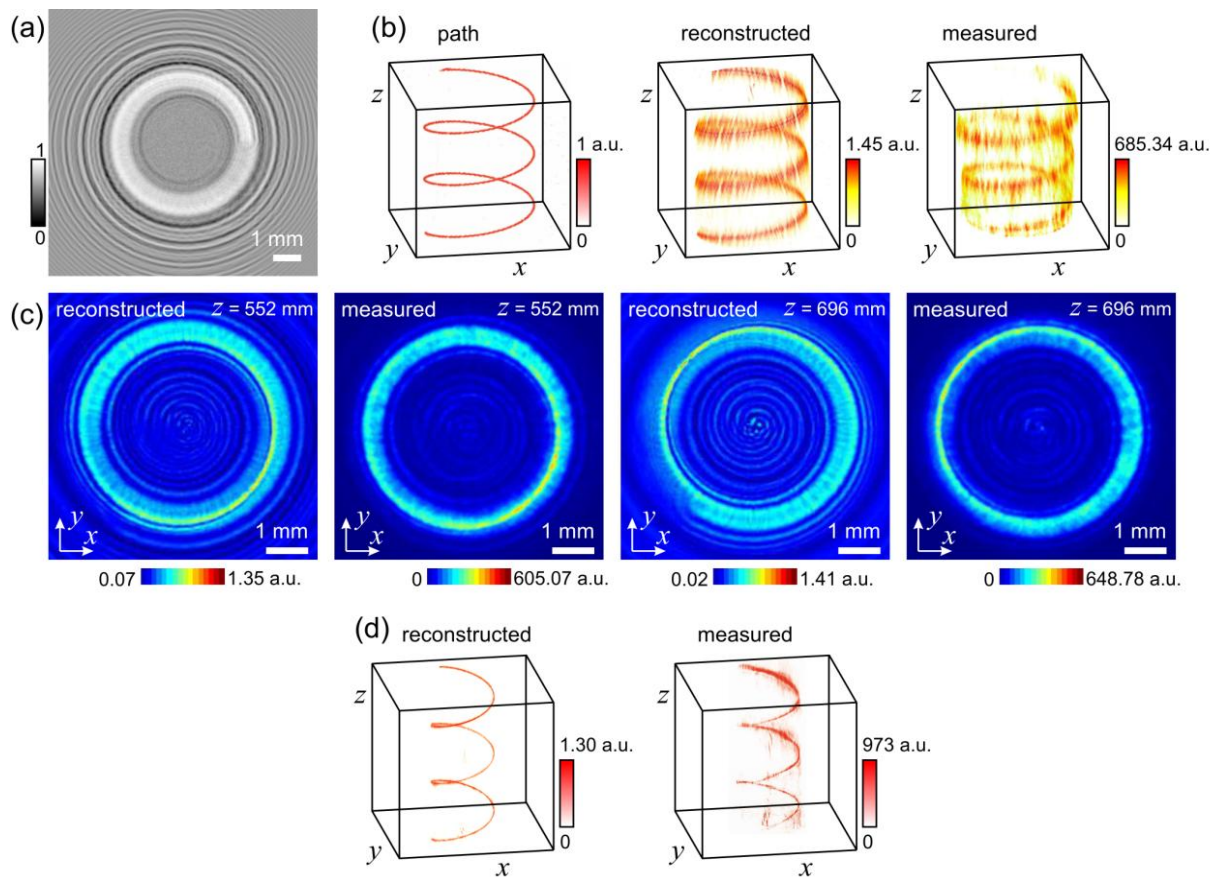

Figure 5. Bending light into a helical shape. (a) Central part of 300 × 300 pixels corresponding to 9.6 mm × 9.6 mm of the simulated diffractor. (b) Three-dimensional representation of the light intensity: the desired intensity path used for simulation of the diffractor (left), the numerically reconstructed intensity distribution by field propagation behind the diffractor (center), and the experimentally measured intensity distribution behind the diffractor (right). (c) Two-dimensional intensity distributions at two different z-positions, $z_1$ = 552 nm and $z_2$ = 696 mm presented in pairs: numerically reconstructed intensity distribution (left) and experimentally measured intensity distribution (right). (d) Three-dimensional representation of the light intensity with optimized sampling parameters: the numerically reconstructed intensity distribution behind the diffractor (left), and the experimentally measured intensity distribution behind the diffractor (right).

**Modulation of intensity with local switching-off**

The intensity can be also modulated in a non-continuous way, for example it can be "switched off" at pre-defined regions for example. The related simulated diffractor is shown in Fig. 6(a) and the created intensity distribution is shown Fig. 6(b), where two parallel beams are joined into one beam, and at three locations the intensity is "switched off". At one location the intensity is "switched off" right after the two beams have merged. It looks as if in the merged beam the intensity is turned off while an increase in intensity would have been expected. An object placed into the region of zero intensity will neither experience light absorption nor scattering and thus, the object will become "invisible". Also, as evident from Fig. 6(b), the intensity re-appear after the region of zero intensity, and the light ray continues its trajectory as if there was no region of zero-intensity. Thus, the achieved effect is analogous to the effect created by utilization of an optical cloak, quote from Prof. John Pendry lecture named "Metamaterials and the Science of Invisibility" at the Imperial College London in 2013: "The function of a cloak is to grab rays of light and steer them away from the object, so you never see it, but also return them to the path which they had before their path was disturbed. So, observer standing here sees what is behind the object and he is unaware of this deviation which has been made."

In the simulation and reconstruction shown in Fig. 6, we sampled at planes $z$ = 100 – 900 mm with steps of 4 mm.

For a three-dimensional intensity modulation providing a situation where the light intensity is locally switched off, the simulation of the diffractor consists of three steps: (1) A complex-valued wavefront in the hologram plane $U_1(X,Y)$ is simulated for the entire optical path as described above by Eq. 7. (2) In the same manner a complex-valued wavefront $U_2(X,Y)$ is simulated for those parts of the optical path that should be excluded from the entire optical path. The reference wave 1 is subtracted from the result giving $U_3(X,Y) = U_2(X,Y) - 1$. (3) The square of the absolute value of the difference $(U_1(X,Y) - U_3(X,Y))$ is calculated and serves as hologram, $H(X,Y) = |U_1(X,Y) - U_3(X,Y)|^2$. The diffractor is calculated as $D(X,Y) = 1 - H'(X,Y)$, where $H'(X,Y)$ is the normalized hologram.

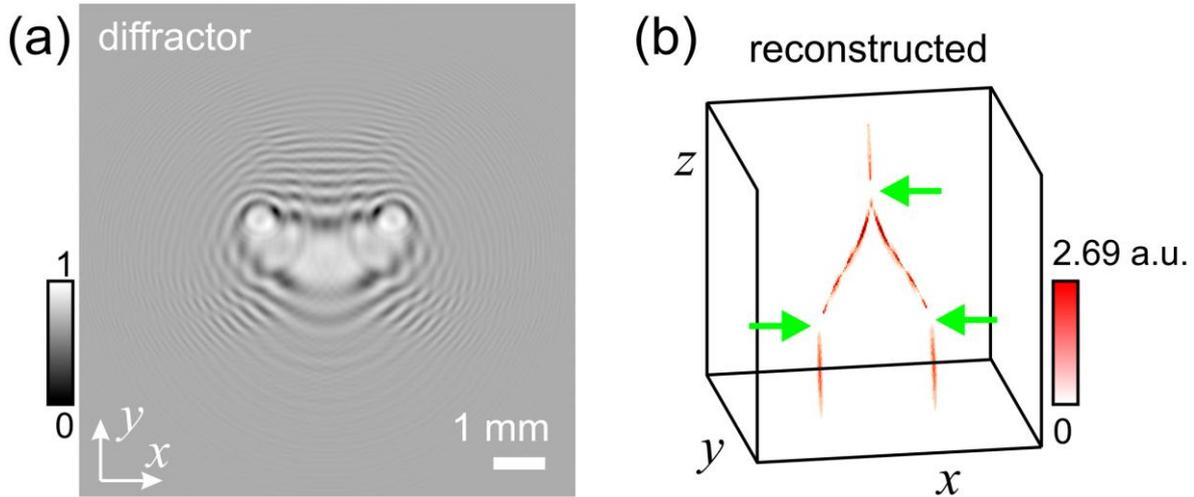

Figure 6. Intensity modulated in such a way that it is switched off at three regions in space. (a) Central part of 300 × 300 pixels corresponding to 9.6 mm × 9.6 mm of the simulated diffractor. (b) Three-dimensional reconstruction showing two parallel beams which are merging into one beam with three predefined regions where the intensity is close to zero, indicated with the green arrows.

**Phase modulation**

Not only three-dimensional shaping of intensity, but also three-dimensional shaping of the phase distribution can be achieved as illustrated by the example in Fig. 7. Here, the variations of the phase should be selected not to exceed π to avoid phase wrapping effects. For a three-dimensional phase modulation, the hologram is simulated as described above with the only difference that the transmission functions $t_i(x,y,z_i)$ are assigned to each plane so that $t_i(x,y,z_i) = 1$ everywhere except at the positions of the point-like objects where $t_i(x=x_i, y=y_i, z_i) = \exp(i\varphi_i)$, with $\varphi_i$ being the introduced (imposed) phase shift. At these positions, the phase of the propagating wave is replaced with $t_i(x=x_i, y=y_i, z_i) = \exp(i\varphi_i)$. Here, the simulated hologram does not need to be inverted for the reconstruction process. The diffractor shown in Fig. 7(a) was simulated with 2000 sampling points in $z$-direction, and the three-dimensional reconstruction shown in Fig. 7(b) was obtained at 800 reconstruction planes in $z$-direction.

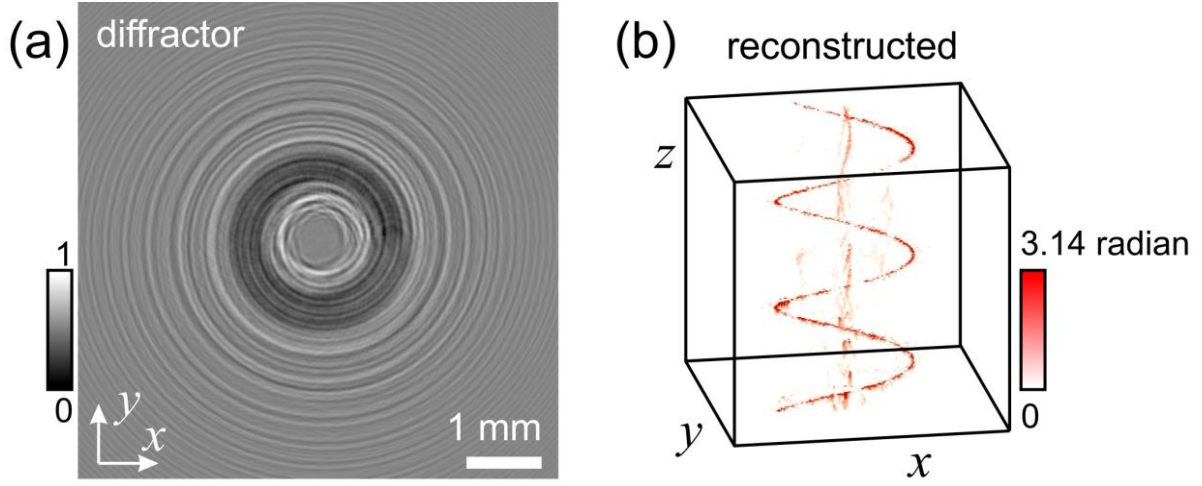

Figure 7. Shaping the phase distribution into a curved trajectory. (a) Central part of 300 × 300 pixels corresponding to 9.6 mm × 9.6 mm of the simulated diffractor. (b) Three-dimensionally reconstructed phase distribution re-creating the spiral form.

**Employing spherical waves**
It can be shown that the optical wavefront does not need to be planar. The fact that holograms recorded with plane wavefront can be reconstructed with divergent wavefront[24] allows the modulation technique to be applied also to a divergent wavefront which can exceed the lateral size of SLM area, as illustrated in Fig. 8. This can be an advantage when compared to the bending Airy beams created with a similar optical setup where typical bending ranges of about 1 mm over a 30 cm propagation distance are achieved[12].

A diffractor which has been created to operate with plane waves, see Fig. 8(a) – (b), can be illuminated with a spherical wave that has its origin at a distance $z_0^{(S)}$ in front of the diffractor. The field behind the diffractor at a distance $z_i^{(S)}$ in plane $\left(x_i^{(S)}, y_i^{(S)}\right)$ is then given by:

$$U_i\left(x_i^{(S)}, y_i^{(S)}, z_i^{(S)}\right) \approx \frac{1}{\lambda z_0^{(S)}} \frac{1}{\lambda z_i^{(S)}} \frac{z_i^{(S)}}{z_0^{(S)}} \iint \exp\left[\frac{i\pi}{\lambda z_0^{(S)}}\left(X^2 + Y^2\right)\right] D(X,Y) \exp\left[-\frac{i\pi}{\lambda z_0^{(S)}}\left((X-x_0)^2 + (Y-y_0)^2\right)\right] \times$$

$$\times \exp\left[\frac{i\pi}{\lambda z_i^{(S)}}\left((x_0 - x_i^{(S)})^2 + (y_0 - y_i^{(S)})^2\right)\right] dX dY dx_0 dy_0 \sim FT\left\{FT^{-1}\left[D(X,Y)\right]\exp\left[-\frac{i\pi}{\lambda z_0^{(S)}}\left(x_0^2 + y_0^2\right)\right]\exp\left[\frac{i\pi}{\lambda z_i^{(S)}}\left(x_0^2 + y_0^2\right)\right]\right\},$$

or 
$$\left|U_i\left(x_i^{(S)}, y_i^{(S)}, z_i^{(S)}\right)\right| \approx \left|FT^{-1}\left\{FT\left[D(X,Y)\right]\exp\left[\frac{i\pi}{\lambda z_0^{(S)}}\left(x_0^2 + y_0^2\right)\right]\exp\left[-\frac{i\pi}{\lambda z_i^{(S)}}\left(x_0^2 + y_0^2\right)\right]\right\}\right|, \tag{10}$$

where we employed the following two-step approach for spherical wave propagation: (1) propagation from the diffractor plane $(X,Y)$ to the spherical wave origin plane $(x_0, y_0)$, and (2) propagation from the spherical wave origin plane $(x_0, y_0)$ to the image plane $\left(x_i^{(S)}, y_i^{(S)}\right)$[28]. The spherical wave term needs to be expressed in digital form:

$$\exp\left[\frac{i\pi}{\lambda}\left(\frac{1}{z_0^{(S)}} - \frac{1}{z_i^{(S)}}\right)\left(x_0^2 + y_0^2\right)\right] = \exp\left[\frac{i\pi\lambda\left(z_0^{(S)}\right)^2}{s^2}\left(\frac{1}{z_0^{(S)}} - \frac{1}{z_i^{(S)}}\right)\left(m^2 + n^2\right)\right], \tag{11}$$

where $s \times s$ is the area size in $(X,Y)$ plane. By comparing Eq. 9 and 11 we obtain:

$$z_i = \left(z_0^{(S)}\right)^2 \left(\frac{1}{z_0^{(S)}} - \frac{1}{z_i^{(S)}}\right). \tag{12}$$

By solving the last equation for $z_i^{(S)}$ we obtain:

$$z_i^{(S)} = z_0^{(S)}\left(1 - \frac{z_i}{z_0^{(S)}}\right)^{-1}, \tag{13}$$

where $z_i^{(S)}$ is the distance at which the wavefront distribution observed under illumination of the diffractor with a spherical wave is analogical to the wavefront distribution observed at distance $z_i$ under illumination of the diffractor with a plane wave but magnified by the factor $M_i = \frac{z_i^{(S)}}{z_0^{(S)}}$, see Fig. 8(c). From Eq. 13 it follows that in order to obtain $z_i^{(S)} > 0$, the condition $z_0^{(S)} > z_i$ must be fulfilled. For example, for a cosine-like intensity path shown in Fig. 8, $z_i$ ranges from 100 to 900 mm. To fulfil the condition $z_0^{(S)} > z_i$ we select $z_0^{(S)} = 1$ m. Using Eq. 13, for $z_1 = 100$ mm we obtain $z_1^{(S)} = 1.111$ m and for $z_2 = 900$ mm we obtain $z_2^{(S)} = 10$ m. The magnification $M_i$ increases from $M_1 = 1.111$ at $z_1 = 100$ mm to $M_2 = 10$ at $z_2 = 900$ mm. The size of the area at $z_i^{(S)}$ is given by the product of the illuminated area on the diffractor and magnification $M_i$, which can be also deduced from simple geometrical considerations schematically shown in Fig. 8(d) – (e). Thus for example the optical wavefront distribution at $z_2^{(S)} = 10$ m is spread over a 10× larger area than the diffractor size. The profile of the obtained optical path is shown in Fig. 8(c) and (e).

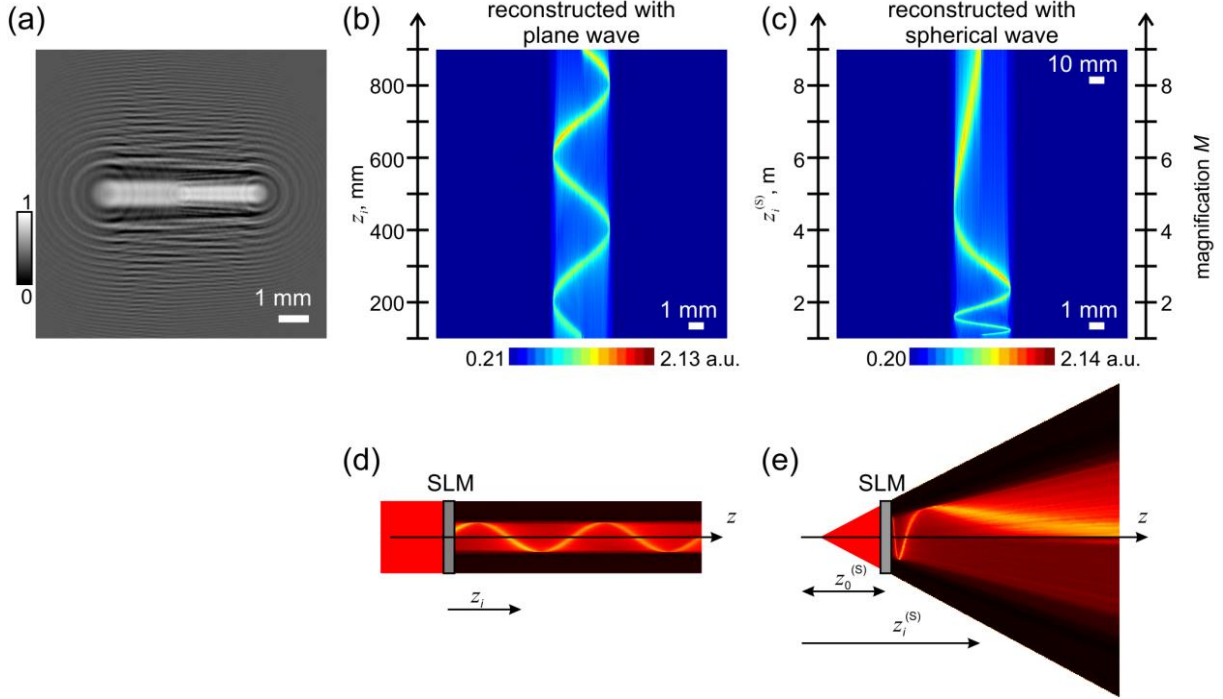

Figure 8. Light bending in a cosine-shape obtained with a spherical wave. (a) The central part of 300 × 300 pixels corresponding to 9.6 mm × 9.6 mm of simulated diffractor. (b) Two-dimensional numerically reconstructed intensity distribution $I(x, y = 0, z)$ behind the diffractor for an incident plane wave. (c) Two-dimensional numerically reconstructed intensity distribution $I(x, y = 0, z)$ behind the diffractor for an incident spherical wave with its origin at 1 m in front of the diffractor. The scale bar at the bottom refers to the wavefront distribution at a distance $z_1^{(S)} = 1.1$ m and the scale bar at the top refers to that at a distance $z_2^{(S)} = 10$ m. (d) and (e) are the geometrical arrangement of the schemes with plane and spherical waves, respectively.

## DISCUSSION

We have demonstrated that the distribution of an optical wavefront can be modulated in such a way that constructive and destructive interference can be achieved in a controllable manner over large distances while the beam is propagating. This in turn leads to a re-distribution of the beam intensity creating an intense optical path of an arbitrary predetermined shape. To create an arbitrary path of intensity, we employed a three-dimensional reconstruction of a contrast-inverted hologram (diffractor) of a sequence of absorbers which are aligned into a predetermined curve. For a smooth appearance of the curve, the distance between the absorbers in $(x, y)$-plane must not exceed the lateral resolution of the optical system, and the distance between the absorbers in $z$-direction must not exceed the axial resolution of the optical system. The spatial resolution at which the path can be created is provided by the resolution intrinsic to in-line holography. The lateral resolution (in $(x, y)$-plane) is given by $R_{\text{lateral}} = \frac{\lambda}{2 \text{N.A.}}$, where N.A. is the numerical aperture of the system and the axial resolution (along z-axis) is given by $R_{\text{axial}} \approx \frac{\lambda}{1.4 \text{ N.A.}^2}$ (see Methods).

The minimum distance from the SLM where the curve of light can start is zero in theory, but in practice the SLM is sandwiched between two polarizers and the curve of light is generated after the second polarizer, that is at a distance of few centimeters behind the SLM. The maximum distance from the SLM where the curve of light can still exist is only given by the intensity of the laser and can range over tens of meters. However, the resolution of the light curve, or its sampling, is given by $R_{\text{lateral}} = \frac{\lambda}{2 \text{N.A.}}$ and $R_{\text{axial}} \approx \frac{\lambda}{1.4 \text{ N.A.}^2}$ (see Methods). It is inversely proportional to N.A. and will decrease with distance accordingly.

The optical wavefront illuminating the diffractor does not need to be planar. We showed that under illumination with a spherical wave, the optical wavefront distribution at $z_i^{(S)} = 10$ m is spread over a 10× larger area than the diffractor size.

This technique of wavefront modulation can be applied to any waves, for example, electron, X-ray waves or terahertz waves provided the diffractor could be scaled down to the nanometer regime. Focusing a wave to a pre-defined path creates a tool with a venue of potential applications such as creating isolated optical knots[29], probing beams focused on certain three-dimensionally distributed areas, laser micro-structuring, optical tweezers[30], shaping intensity of short optical pulses[31], probing selected regions deep in tissues in biomedical research, controllable switching off intensity for optical cloaking.

# METHODS
## Lateral resolution
The lateral resolution (resolution in the plane orthogonal to the optical axis) is given by the Abbe criterion:

$$R_{\text{Abbe}} = \frac{\lambda}{2\text{NA}}, \tag{14}$$

where NA is the numerical aperture of the system:

$$\text{NA} = \sin\theta = \frac{s}{2\sqrt{(s/2)^2 + z_0^2}}, \tag{15}$$

where $\theta$ is the diffraction angle, $s \times s$ is the detector size, and $z_0$ is the distance between the detector and the plane of the focusing point, see Fig. 9(a). At larger $z_0$, with $z_0 \gg s$, the resolution is given by

$$R_{\text{Abbe}} = \frac{\lambda z_0}{s}. \tag{16}$$

Lateral resolution in in-line holography can be derived as following. A hologram of a point-like scatterer, positioned at $(x=0, y=0, z=z_0)$ is given by

$$H(X,Y) = 2 + \exp\left(\frac{i\pi}{\lambda z_0}(X^2 + Y^2)\right) + \exp\left(-\frac{i\pi}{\lambda z_0}(X^2 + Y^2)\right). \tag{17}$$

Assuming that the detector exhibits a square shape of $s \times s$ in size, the reconstruction is given by backward propagation from the hologram plane $(X, Y)$ to the point-like scatterer plane $(x, y)$:

$$o(x,y) = \frac{i}{\lambda z_0}\int_{-s/2}^{s/2}\int_{-s/2}^{s/2} H(X,Y)\exp\left(-\frac{i\pi}{\lambda z_0}\left((x-X)^2 + (y-Y)^2\right)\right)dXdY \sim \frac{i\lambda z_0}{\pi^2 xy}\sin\left(\frac{\pi s x}{\lambda z_0}\right)\sin\left(\frac{\pi s y}{\lambda z_0}\right). \tag{18}$$

The first minima of the sinc-functions are found at

$$x_{\min}, y_{\min} = \frac{\lambda z_0}{s}. \tag{19}$$

Two peaks are resolved when the maximum of one peak coincides with the minimum of the other peak. Thus, the resolution in in-line holography is given by:

$$R_{\text{H}} = \frac{\lambda z_0}{s}, \tag{20}$$

In analogy to the same resolution limit given by the Abbe criterion in Eq. 16.

Lateral resolution in digital in-line holography is derived as following. In digital holography, a hologram is sampled as $s = N \cdot \Delta$, where $N$ is the number of pixels, and $\Delta$ is the pixel size. Equation 20 can be re-written for digital holography as:

$$R_{\text{DH}} = \frac{\lambda z_0}{N\Delta}. \tag{21}$$

## Axial resolution
For the axial resolution (resolution along the optical axis), a diffraction calculation cannot be achieved as it would require the evaluation of the out-of focus reconstruction of a point scatterer which cannot be obtained analytically. The axial resolution is derived by the following analysis. If a point source is radiating energy in all directions thus producing a spherical wave, then the intensity decreases in proportion to the squared distance from the object. The net power which is the intensity integrated over the surface remains constant. A similar consideration can be applied to a wave scattered off a point-like object positioned on the optical axis. Energy conservation demands that the net power behind the absorber will result in constant $P_0$ when the intensity is integrated over the surface of the scattered wave. The reconstructed object intensity is the square of the reconstructed amplitude of the object wave. The reconstructed intensity distribution of the scatterer in the focus plane is spread over a non point-like but finite spot whose diameter is given by the resolution $R_H$. Assuming a constant intensity within the spot, the intensity at the centre of the spot (on the optical axis) is given by

$$\frac{P_0}{\pi R_H^2}. \tag{22}$$

We define the axial resolution $z_R$ as the distance from the in-focus plane, where the reconstructed object intensity is maximal, to the plane where the reconstructed object intensity is reduced to half, see Fig. 9. Thus, at the distance $z_R$, the intensity on the optical axis is given by:

$$\frac{P_0}{2\pi R_H^2}. \tag{23}$$

On the other hand, the intensity at distance $z_R$ is distributed over an area $\pi(s_R/2)^2$, where $s_R$ is the diameter of the spot. Assuming a constant intensity within the spot, the intensity on the optical axis equals to:

$$\frac{P_0}{\pi(s_R/2)^2} = \frac{P_0}{\pi(z_R \sin\theta)^2}. \tag{24}$$

By comparing Eqs. 23 and 24 we obtain:

$$2R_H^2 = (z_R \sin\theta)^2, \tag{25}$$

and extract $z_R$ as

$$R_{\text{axial}} = z_R = \frac{\sqrt{2}R_H}{\sin\theta} = \frac{\lambda}{\sqrt{2}\text{NA}^2} \approx \frac{\lambda}{1.4\text{NA}^2}, \tag{26}$$

where we used Eqs. 14 and 15. The axial resolution given by Eq. 26 differs from the axial resolution obtained in optical microscopy $R_{\text{axial}}^{\text{optical microscopy}} \approx \frac{1.4\lambda}{\text{NA}^2}$.

An example of the axial resolution estimation in an in-line hologram is provided in Fig. 9. A simulated optical hologram of a point-like absorber positioned at a distance of 80 mm from the detector plane is shown in Fig. 9(b); with a wavelength of 500 nm and a hologram size of $2 \times 2$ mm$^2$. This provides a numerical aperture of $\text{NA} = 0.0125$ and an axial resolution $R_{\text{axial}} = 2.26$ mm obtained by Eq. 26. The reconstructed object intensity distribution in the $(x,z)$-plane is shown in Fig. 9(c). From the analysis of the reconstructed object intensity, we obtain the position of the plane where the intensity drops by half at a distance 2.2 mm from the in-focus plane. This distance is very close to the axial resolution obtained by Eq. 26. The object intensity distributions at the in-focus plane and at the plane where the intensity drops by half are shown in Fig. 9(d) and their magnified regions are shown in Fig. 9(e).

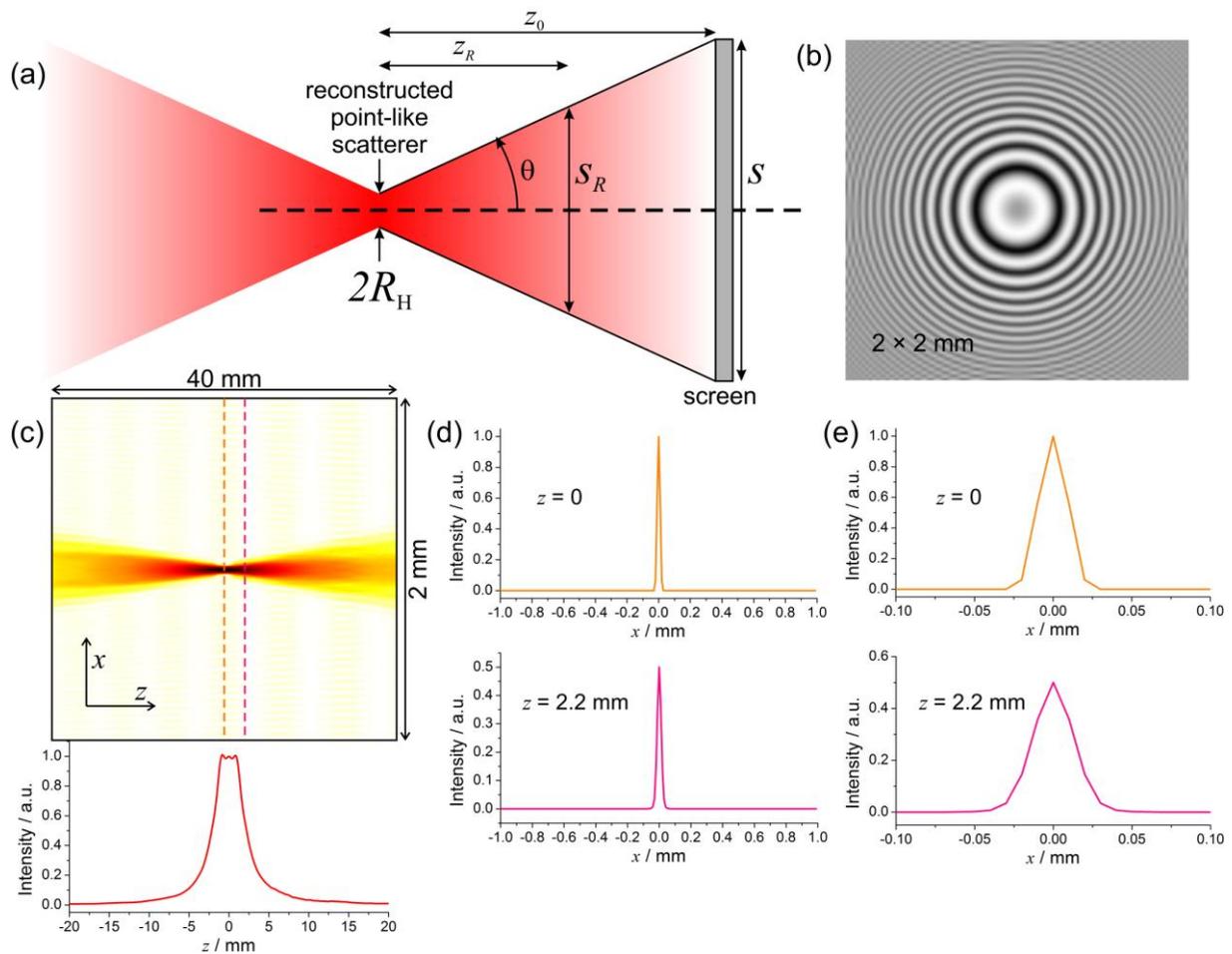

Figure 9. Lateral and axial resolution in the case of in-line holography. (a) Illustration to the symbols. (b) A simulated optical hologram of a point-like absorber positioned at a distance 80 mm from the detector plane, wavelength 500 nm, the hologram size is $2 \times 2$ mm$^2$, the numerical aperture $NA = 0.0125$. The contrast of the hologram is inverted. (c) Intensity distribution $(x, z)$ reconstructed from the inverted hologram. (d) Intensity profiles at $z = 0$ and $z = 2.2$ mm where the intensity drops by half, indicated with dashed lines in (c). (e) Magnified regions of the intensity profiles in (d).

## Acknowledgements
Financial support of the University of Zurich is acknowledged.

## Contributions
T.L. initiated the project, performed the simulations and the experiments, prepared the figures, participated in discussions and contributed to writing the manuscript.
H.-W. F. participated in discussions and contributed to writing the manuscript.